\documentclass{jpsj2}

\def\runtitle{
Relationship between Height and Spacing in Barchan Dunes}
\def\runauthor{Shibata and Taguchi}

\title{%
Verifying relationship between Height and Spacing, in Barchan Dunes simulated
by the Coupled Map Lattice Model
}

\author{%
$^1$Jun Shibata\thanks{jun@phys.chuo-u.ac.jp} and 
$^{1,2}$Y-h. Taguchi\thanks{tag@granular.com}
}

\inst{%
$^1$Department of Physics,Chuo University,
Bunkyo-ku,Kasuga,Tokyo 112-8551 \\
$^2$The Institute of Science and Engineering, Chuo University,
Bunkyo-ku,Kasuga,Tokyo 112-8551
}

\recdate{\today}

\abst{%

We have investigated the relationship between height and spacing
of Barchan dunes which the coupled map lattice model numerically generates.
There is a scaling relation between them and
the values of the scaling exponents agree well with real dunes' values.
The values of these scaling exponents are the same for both
steady states and transient states.
}

\kword{%
Barchan dune, coupled map lattice, scaling exponents
}

\begin{document}
\sloppy
\maketitle

\section{Introduction}
There are many serious problems all over the world. One of them is how to control the behavior of dunes.
So the problem of controlling dunes should be solved and it has recently begun to be researched quantitatively by a number of physicists.\par
There is an experimental observation; height $H$ and spacing $L$ of dunes have a relation
\begin{equation}
L\sim H^{a},
\end{equation}
where $a=0.58\sim 1.92$\cite{lancaster}.
Since it is a non-trivial observation, some theoretical analysis is necessary.
Unfortunately there are no theoretical approaches for this relation. 
Instead we have a powerful method; numerical analysis using computer. 
In this research, we would like to investigate the validity of this relation
by computer simulations, "Coupled Map Lattice (CML) model". 
It was developed by Kaneko\cite{kaneko}, and was applied to researches of
dunes by Nishimori and Ouchi \cite{nishimori1,nishimori2}. \par
A crescent-shaped barchan dune  is frequently observed in a desert. 
We analyzed barchan dunes by the numerical simulation with CML model.
 In this paper, "dune" means "barchan dune".\par
In the next section, we will discuss the relationship
between height and spacing of real dunes. 
In the third section, we will show some results by computer simulation.
Some discussions will be included in the fourth section.

\section{Experimental Law and Assumption}
When there are $N$ dunes, the $n$th dune is supposed to have height $h_n$ and mass $m_n$. It is expected that they have similar shapes with each other, because granular matter has an angle of repose. 
However, if the sand particles are blown by the wind,
smaller dune 
may lose more sand than the larger ones.
They will break the similarity of shapes. Therefore, the mass of a dune
is not proportional to 3rd power of height.
Sauermann {\it et al.}\cite{sauermann} show that
\begin{equation}
m_n= \gamma h_n^{2.4},
\end{equation}
where, $\gamma $ is a constant number, a total sand mass $M_{total}$ is given by
\begin{equation}
M_{total}=\gamma \sum _{n}^{N}h_{n}^{2.4}.
\end{equation}
It is assumed that this $M_{total}$ is shared by individual dunes.\par
Barchan dunes are formed when $M_{total}$ is relatively small and the
wind direction does not fluctuate. 
Thus we can choose two axes. One is parallel to the wind direction and the other is perpendicular to the same one.
Along one of these axes, the distance between dunes can be measured as
$L_{\| }$ or $L_{\bot }$. $L_{\| }$ is a spacing of dunes along the direction parallel 
to the wind direction, and $L_{\bot}$ is a spacing of dunes along the
direction perpendicular to the  wind direction.
 Here, we know the area $S_n$ which $n$th dune occupies, 
\begin{equation}
S_n\sim L_{\| }L_{\bot }.
\end{equation}
There is another expression of $S_{n}$ using the area $S_{total}$ occupied by N dunes.
\begin{equation}
S_{n}=\frac{S_{total}}{N}=\frac{S_{total}}{\frac{M_{total}}{\langle m_n\rangle }}=\frac{l^2}{\frac{M_{total}}{\gamma \langle h_{n}\rangle ^{2.4}}}=\gamma \frac{l^2}{M_{total}}\langle h_{n}\rangle ^{2.4},
\label{eq5}
\end{equation}
where $S_{total}$ is an area of this system, $l$ is size of the observed area.\par
We assume that two relations,
\begin{equation}
L_{\| }\sim \langle h_n\rangle ^{\alpha },
\end{equation}
\begin{equation}
L_{\bot }\sim \langle h_n\rangle ^{\beta },
\end{equation}
the occupied area by the $n$th dune is
\begin{equation} 
S_{n}\propto L_{\| }L_{\bot } \propto \langle h_{n}\rangle ^{\alpha}
\langle h_{n}\rangle ^{\beta}=\langle h_{n}\rangle ^{\alpha+\beta}.
\label{eq8}
\end{equation}
(\ref{eq5}) is compared with (\ref{eq8}), the exponent of (\ref{eq8}) is expected as
\begin{equation}
\alpha +\beta =2.4 .
\end{equation}

\section{Numerical experiments and Results}
We used CML model which was arranged by Nishimori and Ouchi
\cite{nishimori1, nishimori2} in order
to apply to computer simulation. At first, we prepare a 
$1000\times1000$ lattice and give an uniform random number $[0,X]$ to all sites. 
These random numbers correspond to the height for each site; 
\begin{equation}
h(i,j) \in [0,X],
\end{equation}
$X$ is the  parameter which controls the total amount of the sand,
 and $h(i,j)$ is an amount of sand at a
site $(i,j)$.\par
Saltation and creep are important when considering a dynamics of dune. Saltation is a flying of sand by the wind. Creep is that sand
rolls and falls with gravity.
Although they had better to be considered exactly,
they are approximated for simplicity.\par
Flight distance $L_{s}$ and the amount of the sand which flies 
by saltation $q_{s}$ have the relationship with $\Delta h$,
which  is a difference of height along the leeward direction\cite{keisu},
\begin{equation}
q_{s}=-\tanh (\Delta h)+1.3,
\end{equation}
\begin{equation}
L_{s}=\tanh (\Delta h)+1.
\end{equation}
We consider that wind is blowing to the positive direction of the $i$ axis,
\begin{equation}
\Delta h(i,j) =h(i+1,j)-h(i,j).
\label{eq13}
\end{equation}
Thus saltation is decided by only landform. \par
Next, we consider creep, which is regarded as "diffusion of sand ". 
We introduce an arbitrary diffusion constant $D$, 
and the amount of diffusion of sand $q_{c}$ is described as
\begin{equation}
q_c=D\Delta h(i\pm 1,j\pm 1) ,
\label{eq14}
\end{equation}
using (\ref{eq13}) and (\ref{eq14}), we get
\begin{equation}
h(i\pm 1,j\pm 1) \to h(i\pm 1,j\pm 1)+q_c,
\end{equation}
\begin{equation}
h(i,j) \to h(i,j)-q_c .
\end{equation}
But, we must not forget "critical angle" of sand; it is an "angle of repose" of sand. 
According to the observation of sand in a hourglass, it is about $34^{\circ }$.
The creep does not occur, if an inclination is equal to $34^{\circ }$ 
or is less than that.
Since $\tan 34^{\circ }\sim 0.67$, only when
\begin{equation}
h(i\pm 1,j\pm 1)-h(i,j) > 0.67 
\end{equation}
creep occurs.\par
We repeat the whole process.  One step contains all of 
these. With this CML model,
a description of dynamics of dunes becomes  much easier  than a real situation.
Nishimori and Ouchi \cite{nishimori1,nishimori2} have already reported that
their model reproduced dune patterns qualitatively.
We are interested in how similar it is
quantitatively to the real dunes. 

\subsection{Scaling Relation in Steady States}

Dunes get to the steady state after some transient period. 
In order to realize this stationary state,
we have iterated whole processes over sufficiently long
period. In this case, it is 3,000 steps.

We will discuss the average height of dunes. 
We do not employ  simple averages over individual dunes.
Instead we use the total height of sand, $\langle H_{st}\rangle $ which can be defined as
\begin{equation}
\langle H_{st}\rangle  \equiv H_{st}/X = \sum _{i,j} h_{st}(i,j)^2/X.
\end{equation}

\begin{figure}[h]
  \begin{center}
    \includegraphics[width=80mm]{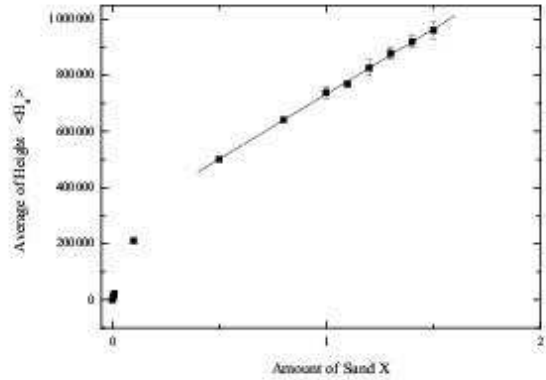}
  \end{center}
  \caption{Relation between mass of sand of average of height of saturated dunes}
  \label{fig:linear.eps}
\end{figure}

Since it is assumed that system size is fixed, $\langle H_{st} \rangle$ is
proportional to the average of height (see Appendix). After some period, 
the system becomes steady states. $h_{st}(i,j)$ expresses the quantity of the sand in a site $(i, j)$ at that time. Figure \ref{fig:linear.eps} shows the
relation between $X$ and a quantity $\langle H_{st}\rangle $ which is 
proportional to an average height of these steady dunes. 
There is  a linear relation between $X$ and $\langle H_{st}\rangle $ as
\begin{equation}
\langle H_{st}\rangle /10^5= (4.64\pm 0.07)X+(2.70\pm 0.07),
\end{equation} 
where CML model generates dune-like patterns only when
this relation stands.
Therefore this equation enables us to judge whether
the system converges to the steady state or not.\par
We examine both the quantity $\langle H_{st}\rangle$ which is proportional to height  and the spacing\cite{L} $L_{st \| }$ which is along the
direction parallel to the wind direction (see Fig.2), and get
\begin{equation}
L_{st \|}\sim \langle H_{st}\rangle ^{0.8\pm 0.5}.
\label{eq19}
\end{equation}

\begin{figure}[h]
\begin{minipage}{8cm}
  \begin{center}
    \includegraphics[width=80mm]{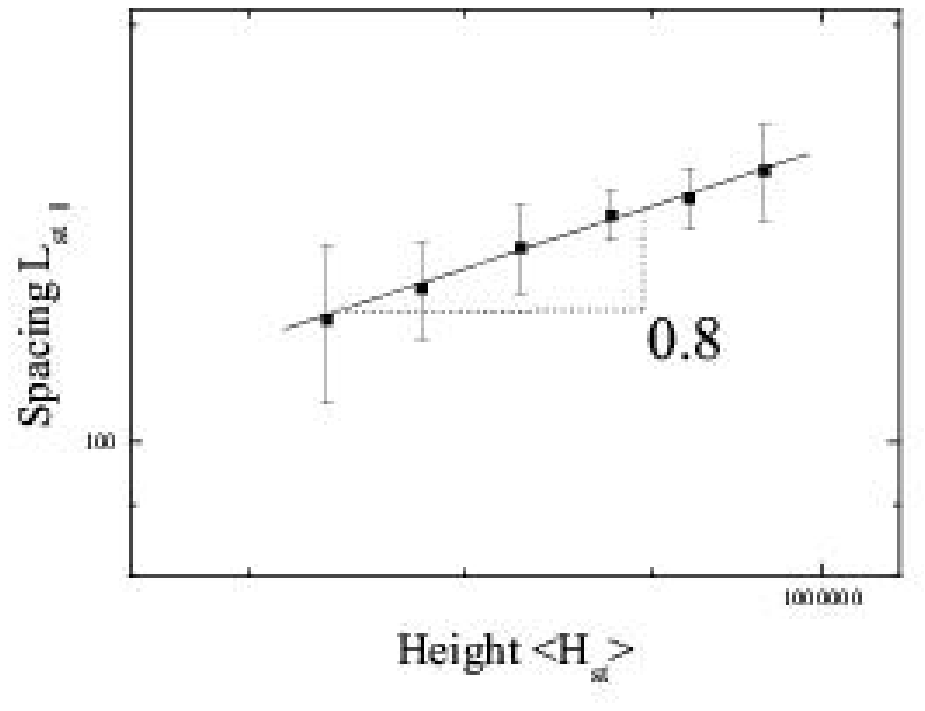}
  \end{center}
  \label{fig:Gyoko.EPS}
\end{minipage}
\begin{minipage}{8cm}
  \begin{center}
    \includegraphics[width=80mm]{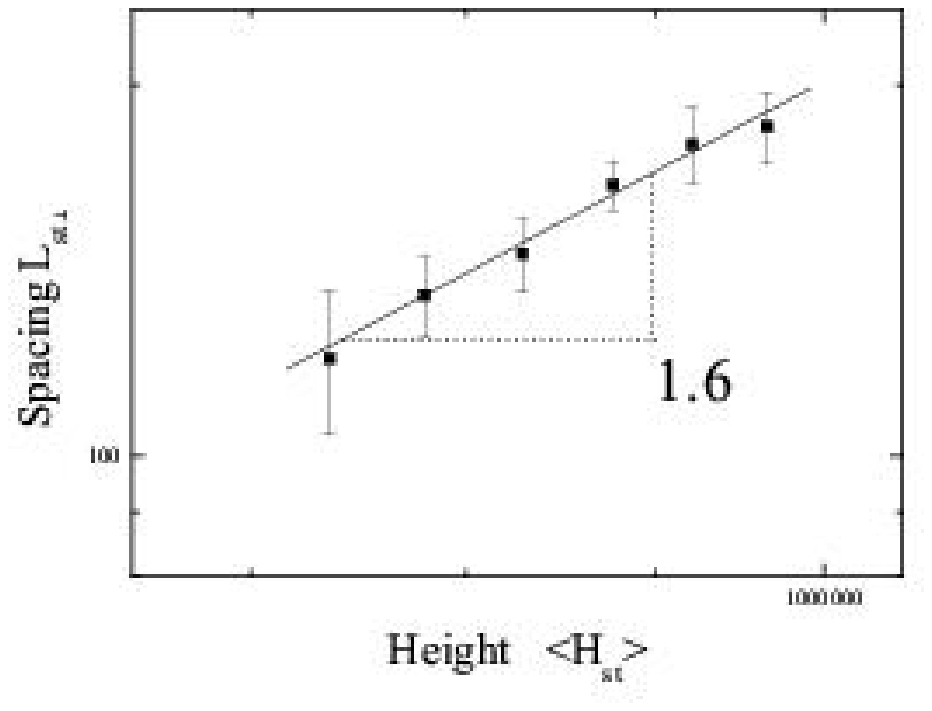}
  \end{center}
\end{minipage}
  \caption{Relation between an average height and a spacing of dunes (left;parallel to 
  the wind direction, right;vertical to the wind direction)}
  \label{fig:Gtate.EPS}
\end{figure}

On the other hand, a quantity $\langle H_{st}\rangle $ has the relation with the spacing 
$L_{st \bot}$ along the direction perpendicular to the wind direction,
\begin{equation}
L_{st \bot}\sim \langle H_{st}\rangle ^{1.6\pm 0.1}.
\label{eq20}
\end{equation} 
Substituting the exponents of (\ref{eq19}) and (\ref{eq20}) into the relation 
(\ref{eq8}), we get
\begin{equation}
S_n\propto \langle H_{st}\rangle  ^{2.4\pm 0.5}.
\end{equation}
This result is consistent with the results obtained in real dunes as
described in the section 2.

\subsection{Scaling Relation in Transient States}

Next we consider the scaling relation in transient states. 
Here, we employ two methods for this investigation.
One is for a fixed time and another is time series scaling.

\subsubsection{One Fixed Time}

At first, we consider a certain fixed time in a transient state
$(t=2,000$ steps$)$, and investigate the relation between the spacing of dunes $L_{fx \| }$ which is along the direction parallel to wind direction and a quantity $\langle H_{fx}\rangle $ (see Fig.3).
A quantity $\langle H_{fx}\rangle $ is defined as,
\begin{equation}
\langle H_{fx}\rangle  \equiv \sum _{i,j} h_{fx}(i,j)^2/X.
\end{equation}
Since system-size is fixed, $\langle H_{fx} \rangle$ is proportional to the average of height at one fixed time in transient states.
$h_{fx}(i,j)$ is the quantity of the sand in a site $(i, j)$ at that time. And we get,
\begin{equation}
L_{fx \| }\sim \langle H_{fx}\rangle ^{0.8\pm 0.1}.
\label{eq22}
\end{equation}

\begin{figure}[h]
\begin{minipage}{8cm}
  \begin{center}
    \includegraphics[width=80mm]{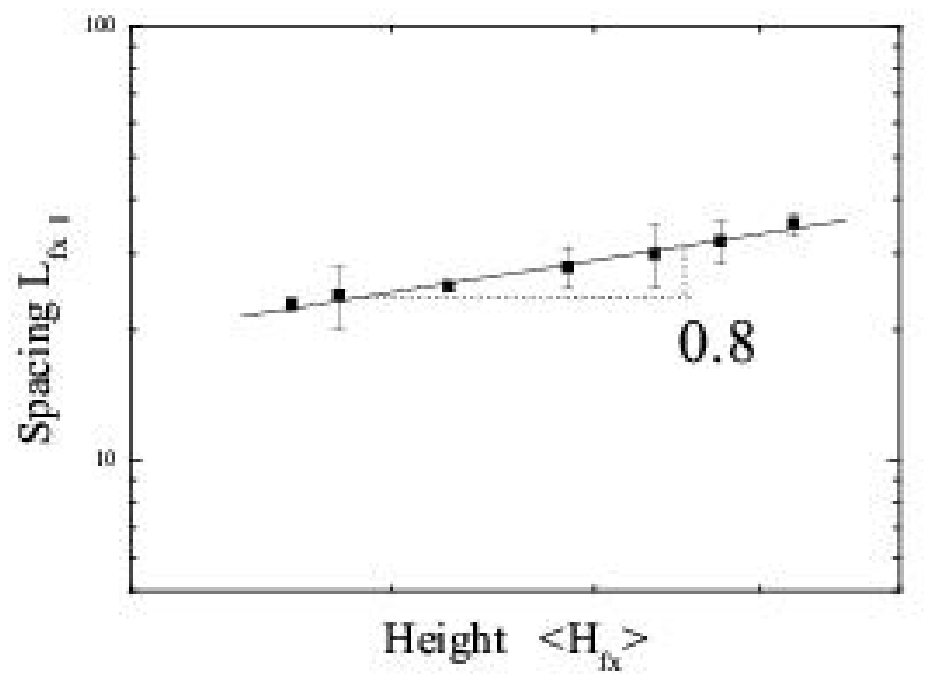}
  \end{center}
  \label{fig:Gyoko-s.EPS}
\end{minipage}
\begin{minipage}{8cm}
  \begin{center}
    \includegraphics[width=80mm]{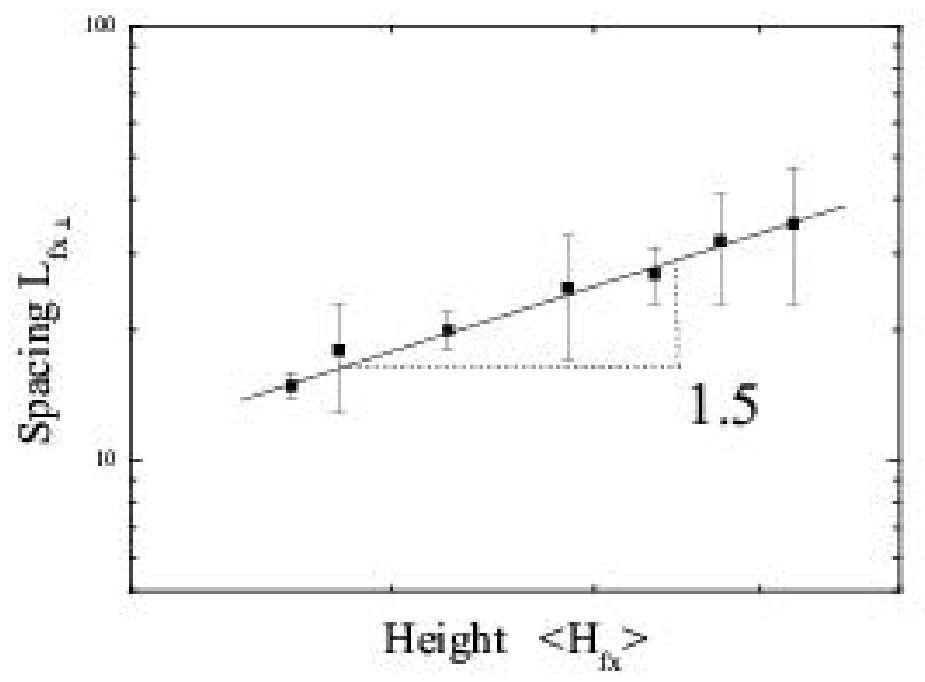}
  \end{center}
\end{minipage}
  \caption{Relation an average height and spacing of dunes (left;parallel to
the wind direction,right;vertical to the wind direction)}
  \label{fig:Gtate-s.EPS}
\end{figure}
Next, we compute spacing $L_{fx \perp }$ which is along the direction perpendicular 
to the wind direction, and get
\begin{equation}
L_{fx \perp }\sim \langle H_{fx}\rangle ^{1.5\pm 0.3}.
\label{eq23}
\end{equation}
\par
Substituting the exponents of (\ref{eq22}) and (\ref{eq23}) into the relation (\ref{eq8}), we get
\begin{equation}
\alpha +\beta= 2.3\pm 0.3,
\end{equation}
therefore,
\begin{equation}
S_n\propto \langle H_{fx}\rangle ^{2.3\pm 0.3}.
\end{equation}
This result also agrees to the argument in Sec. 2.

\subsubsection{Time Series Scaling}
The transitional time period varies with 
the initial quantity of sand.
During this period, state of dunes continues to change.
Next, we analyze dunes in transient states, by time series scaling method. \par
The total sand height in a certain time $t$ in transient states is set to $H_{total}(t)$. 
And the quantity $H(t)$ which is proportional to a height of dunes in this states can be defined as
\begin{equation}
H(t) \equiv H_{total}(t)/X= \sum_{i,j} h_{(i,j)}(t)^2/X.
\end{equation}
Here, $h_{(i,j)}(t)$ expresses the height a site $(i,j)$ has in a certain time $t$. And $X$ is a parameter which controls initial quantity of sand.
We normalize variables $H(t)$ and $L$ as
\begin{equation}
(t,H(t)) \to (\frac{t}{\langle H_{st}\rangle ^\delta}, \frac{H(t)}{\langle H_{st}\rangle }) ,
\end{equation}
\begin{equation}
(t,L(t))\to  (\frac{t}{L_{st}^{\delta '}} ,\frac{L(t)}{L_{st}} ).
\end{equation}

Where $\delta ,\delta '$ are scaling indices, and $L_{st }$ means a spacing of dunes in a steady states.
We assume that a scaling relation between time $t$ and height $H(t)$ as follows,
\begin{equation}
\frac{H(t)}{\langle H_{st}\rangle } =f(\frac{t}{\langle H_{st}\rangle ^\delta  } )\sim (\frac{t}{\langle H_{st}\rangle ^\delta } )^{\epsilon  }\hspace{15mm}(t\ll \langle H_{st}\rangle ^{\delta}),
\label{eq29}
\end{equation}
and also  the same scaling relation between $t$ and $L(t)$ is  
assumed
\begin{equation}
\frac{L(t)}{L_{st}} =f(\frac{t}{L_{st}^{\delta '}} )\sim (\frac{t}{L_{st}^{\delta '}} )^{\epsilon '}\hspace{15mm}(t\ll L_{st}^{\delta '}),
\label{eq30}
\end{equation}
removing $t$ from (\ref{eq29}) and (\ref{eq30}), we get
\begin{equation}
\frac{H(t)}{L(t)^{\frac{\epsilon }{\epsilon '}}} \sim \frac{\langle H_{st}\rangle }{L_{st}^{\frac{\epsilon }{\epsilon '}}}(\frac{L_{st}^{\delta '}}{\langle H_{st}\rangle  ^{\delta }})^{\epsilon }.
\label{eq31}
\end{equation}
The right hand side of (\ref{eq31}) is constant. Therefore, the relation $H(t)$ and $L(t)$ is given by
\begin{equation}
H(t)\sim L(t)^{\frac{\epsilon }{\epsilon '}} \longrightarrow L(t)\sim  H(t)^{\frac{\epsilon '}{\epsilon }}.
\end{equation}
The parameter for the spacing along the parallel direction to the wind
direction is
\begin{equation}
\alpha =\frac{\epsilon_\parallel}{\epsilon},
\end{equation}
and for the spacing along the perpendicular direction to the wind direction is 
\begin{equation}
\beta =\frac{\epsilon_\perp}{\epsilon}.
\end{equation}\\

\begin{figure}[h]
  \begin{center}
    \includegraphics[width=80mm]{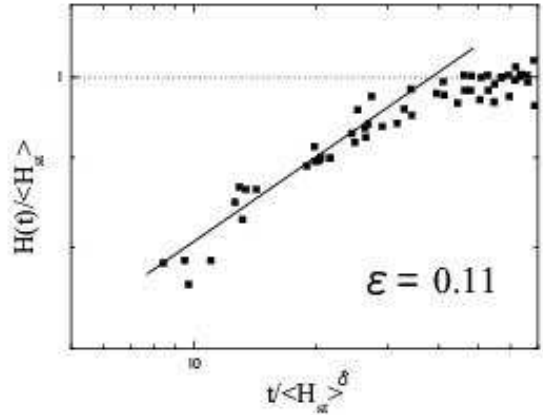}
  \end{center}
  \caption{Time series scaling of the height of dunes}
  \label{fig:Hscaling.EPS}
\end{figure}

\begin{figure}[h]
\begin{minipage}{8cm}
  \begin{center}
    \includegraphics[width=80mm]{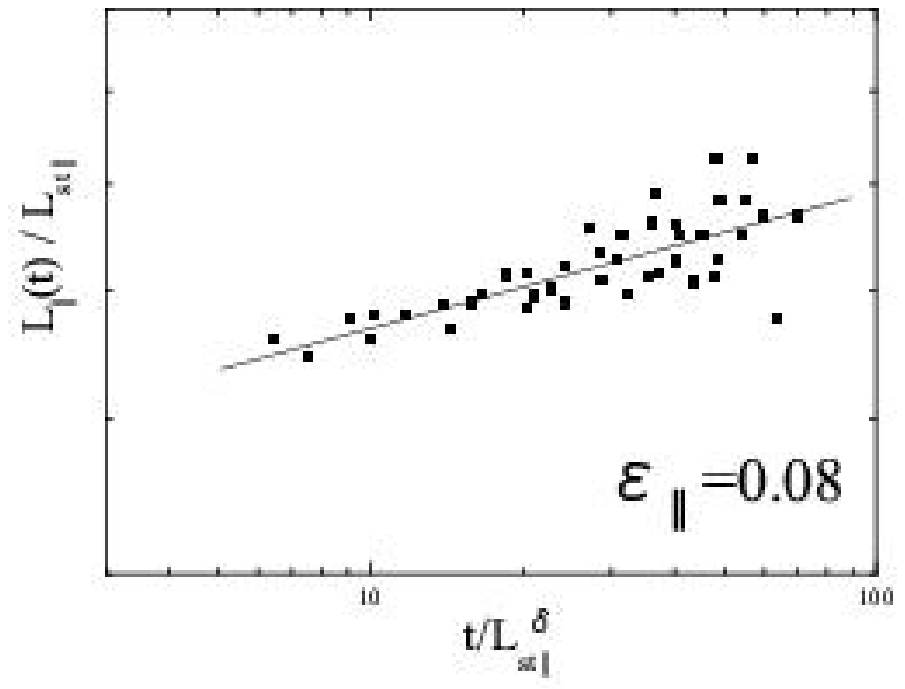}
  \end{center}
  \label{fig:LSyoko.EPS}
\end{minipage}
\begin{minipage}{8cm}
  \begin{center}
    \includegraphics[width=80mm]{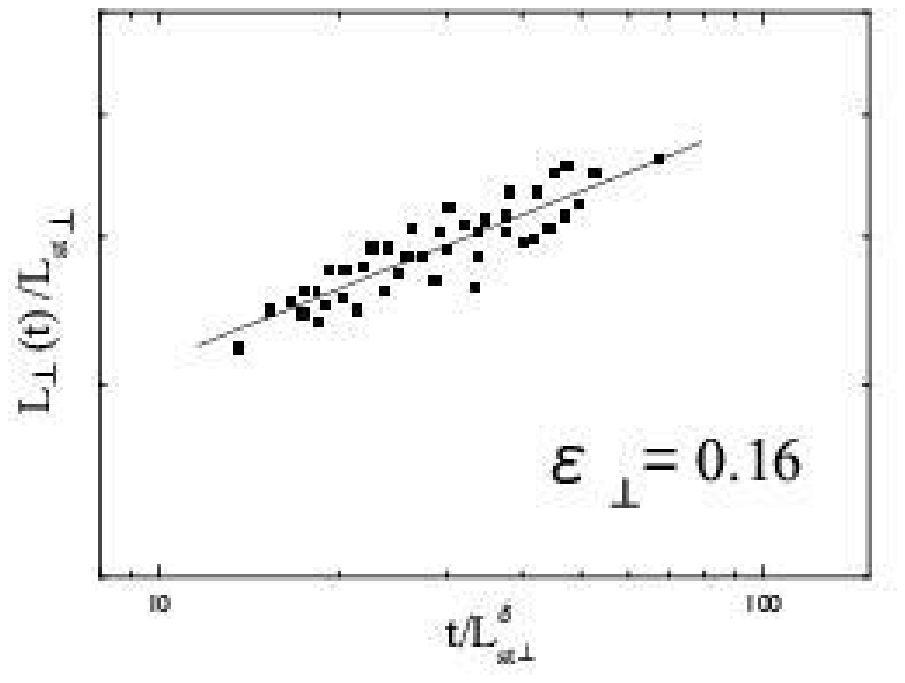}
  \end{center}
  \label{fig:LState.EPS}
\end{minipage}
\caption{
Time series scaling of the spacing of dunes(left: parallel to the wind direction.
right:vertical to the wind direction).  $L_{st \parallel}$
 $L_{st \perp}$ is a spacing along direction parallel to the wind
direction in steady states.}

\end{figure}

$\epsilon_\parallel(\epsilon_\perp)$ is a parameter $\epsilon '$ 
when a parallel
(perpendicular) direction to the  wind direction is considered.
In our simulations, three exponents ($\epsilon, \epsilon_\parallel, \epsilon_\perp$) are (see Fig. 4 and Fig.5)\\
\begin{equation}
\epsilon =0.11 \pm 0.02  
\end{equation}
\begin{equation}
\epsilon_\parallel = 0.08 \pm 0.01
\end{equation}
\begin{equation}
\epsilon_\perp=0.16 \pm 0.01 . 
\end{equation}
Here, we note that $\alpha$ ($\beta$) is the exponent for parallel
(perpendicular) direction to the wind direction and it is estimated as follows,
\begin{equation}
\alpha =\frac{\epsilon_\parallel}{\epsilon }=\frac{0.08 \pm 0.01}{0.11 \pm 0.02} \to 0.7 \pm 0.2 ,
\end{equation}
\begin{equation}
\beta =\frac{\epsilon_\perp}{\epsilon }=\frac{0.16 \pm 0.01}{0.11 \pm 0.02} \to 1.5 \pm 0.4 .
\end{equation}
Here we again give the value
\begin{equation}
S_n \propto H(t)^{2.2\pm 0.4}.
\end{equation}
Again, this value does not disagree with the values obtained previously.

\section{Discussions}

We have estimated the exponent $\alpha +\beta \cong 2.4 $ using
several methods. 
This is the first estimation using numerical simulation. 
These values are consistent with the value by the argument developed from  of
Sauermann's observation \cite{sauermann} in Sec. 2, as shown in Table.1. 
\begin{table}[h]
 \begin{center}
  \begin{tabular}{cccc}
    \hline
     state of dunes & parameter\\
    \hline
     value from observation & $2.4$ \\
     \hline
     steady states & $2.4 \pm 0.5$ \\
     \hline
     transient states (fixed time) $2.3 \pm 0.3$ \\
     \hline
     transient states (scaling)  $2.2 \pm 0.4$ \\
     \hline
  \end{tabular}
 \end{center}
   \caption{Summary of this study}
\end{table}

In the present paper, 
we have discussed a quantity which is proportional to an average height and spacing of dunes. 
All our results agree with Lancaster's observations\cite{lancaster}. 
The CML model approximates dynamics of dunes. 
Pay attention to the fact that 
we did not treat exactly relationship between sand and wind.
But as you saw, our results agreed well with the real dunes.
Thus, exact discussions about the relationship between wind and sand, for example
a sand flux and a distance of flight, may  not be so important. 
A sand flux and a distance of sand flight 
is designed by how wind blows over a dune,
and how wind blows over a dune is designed by the landform.
To tell the truth, even if we know about only landform,
we can guess sand flux $q_s$ and distance of sand flight $L_s$.
Similarly, creep was treated as a diffusion of sand.
We did not consider repellent force and so on which must be considered when we
deal with real sand creep. \par
Thus, we can consider that exact theory about wind velocity, 
sand flux and so on is not so necessary. 
There seems to exist an universality class
about the relationship among landform, sand flux and
distance of sand flight.
Thus the exact creep theory can be replaced with the diffusion theory.
Now we suggest that CML model is suitable model when studying an average height and a spacing of dunes. \par

\begin{center}
{\large \textrm{\textbf{Acknowledgements}}}\\
\end{center}
We would like to thank Prof. H. Nishimori for his great valuable comments and discussions. 
And we thank Mr. T. Nakashima for his careful reading for this paper.

\appendix
\section{Definition of average height}
Consider two differently-shaped dunes with the equal amount of sand.
The following figures show the
cross section along leeward direction.
For simplicity, we assume a simple sinusoidal shape for each dune.
\begin{figure}[h]
  \begin{center}
    \includegraphics[keepaspectratio=true,height=50mm]{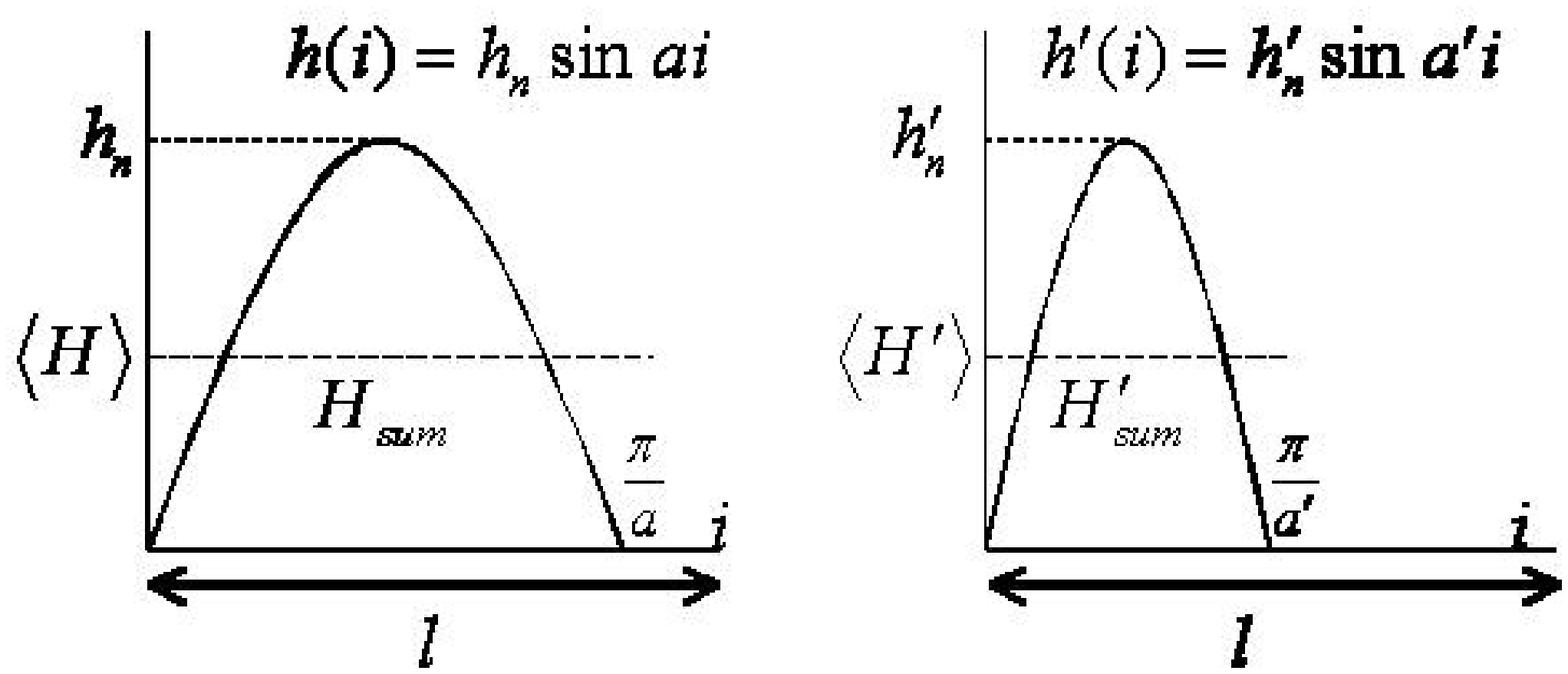}
  \end{center}
\end{figure}
The total height $H_{sum}$ and $H'_{sum}$ are 
\begin{equation}
H_{sum} \equiv \int^{l}_{0}h(i)\mbox{d}i=h_n\int^{\frac{\pi}{a}}_{0}\sin
ai\mbox{d}i=\frac{2h_n}{a},
\end{equation}
and
\begin{equation}
H_{sum}' \equiv \int^{l}_{0}h(i)\mbox{d}i=h_n'\int^{\frac{\pi}{a'}}_{0}\sin
a'i\mbox{d}i=\frac{2h_n'}{a'}.
\end{equation}
These integrated values are equal because the amount of sand is equal. And following relations are realized,
\begin{equation}
\frac{2h_n}{a}=\frac{2h_n'}{a'},
\end{equation}
and
\begin{equation}
h_n:a=h_n':a'.
\label{furoku}
\end{equation}
The average height of dunes $\langle H \rangle $ and $\langle H' \rangle $ becomes
 equal to the integrated value divided  by system size,
\begin{equation}
\langle H\rangle = \frac{2h_n}{al},
\end{equation}
and
\begin{equation}
\langle H'\rangle = \frac{2h_n'}{a'l}.
\end{equation}
When (\ref{furoku}) is considered, $\langle H \rangle = \langle H' \rangle $ is realized and
it turns out that this calculation cannot tell us the height of dunes, $h_n$
or $h_n'$.
When the value of each site is squared, we get the integrations
\begin{equation}
H_{st} \equiv \int^{l}_{0}h(i)^2 \mbox{d}i =h_n^2 \int^{\frac{\pi}{a}}_{0}
\sin^2 ai \mbox{d}i =\frac{h_n^2\pi}{2a},
\label{h_st}
\end{equation}
and
\begin{equation}
H'_{st}\equiv \int^{l}_{0}h(i)^2 \mbox{d}i ={h_n'}^2\int^{\frac{\pi}{a'}}_{0}
\sin^2 a'i \mbox{d}i =\frac{{h_n'}^2\pi}{2a'}.
\label{hp_st}
\end{equation}
Since the average squared-height of dunes $\overline{H_{st}}$ and
$\overline{H'_{st}}$ become equal to the integrated value divided by system
size, next relations are realized,
\begin{equation}
\overline{H_{st}} = \frac{h_n^2\pi}{2al},
\end{equation}
and
\begin{equation}
\overline{H'_{st}} = \frac{{h_n'}^2\pi}{2a'l}.
\end{equation}
When (\ref{furoku}) is considered,
\begin{equation}
\overline{H_{st}}:\overline{H'_{st}} =h_n:h_n'
\label{oH_st}
\end{equation}
is concluded. 
On the other hand, it is also that the ratio of these integration values is,
\begin{equation}
H_{st}:H'_{st}=h_n:{h_n}'.
\label{ratio_H_st}
\end{equation}
From (\ref{h_st}),(\ref{hp_st}), (\ref{oH_st}) and  (\ref{ratio_H_st}), 
\begin{equation}
\overline{H_{st}}:\overline{H'_{st}}=H_{st}:H'_{st}=\int^{l}_{0}h(i)^2\mbox{d}i:\int^{l}_{0}{h(i)'}^2\mbox{d}i.
\end{equation}
Thus the  ratio of  average heights of dunes can be estimated by squared quantity of each site. \par
If we define quantity $\langle H_{st} \rangle$ 
\begin{equation}
\langle H_{st}  \rangle = \frac{H_{st}}{X},
\end{equation}
and consider that  total sand mass is  the product of   $X$ and
the system size $\ell$,
\begin{equation}
H_{sum} = X \ell.
\end{equation}
we get 
\begin{equation}
\langle H_{st} \rangle = \frac{H_{st}}{H_{sum}} \ell \simeq h_n \ell.
\end{equation}
If $\ell$ is fixed, we can use $\langle H_{st} \rangle$ as the quantity proportional to
$h_n$.

\begin{thebibliography}{99}
\bibitem{L}$L_{st \| }$ is estimated from the peak wave number of spatial Fourier
spectrum.
\end{thebibliography}
\end{document}